\documentclass[10pt,twocolumn,letterpaper]{article}

\usepackage{cvpr}
\usepackage{times}
\usepackage{epsfig}
\usepackage{graphicx}
\usepackage{amsmath}
\usepackage{amssymb}

\usepackage[pagebackref=true,breaklinks=true,letterpaper=true,colorlinks,bookmarks=false]{hyperref}

\def\I{\mathbf{I}}
\def\D{\mathbf{D}}
\def\A{\mathbf{A}}
\def\O{\mathbf{O}}
\def\Yl{\mathbf{Y_l}}
\def\Yr{\mathbf{Y_r}}
\def\M{\mathbf{M}}

\def\R{\mathbb{R}}

\def\ie{i.e\onedot}

\cvprfinalcopy % *** Uncomment this line for the final submission

\def\cvprPaperID{12} % *** Enter the WSS Paper ID here

\ifcvprfinal\fi

\begin{document}

\title{Depth Infused Binaural Audio Generation using Hierarchical Cross-Modal Attention}

\author{
Kranti Kumar Parida\textsuperscript{1} \hspace{2em} 
Siddharth Srivastava\textsuperscript{2} \hspace{2em} 
Neeraj Matiyali \textsuperscript{1} \hspace{2em} 
Gaurav Sharma\textsuperscript{1,3} \vspace{0.5em} \\
\textsuperscript{1 }IIT Kanpur \hspace{1.5em}
\textsuperscript{2 }CDAC Noida \hspace{1.5em}
\textsuperscript{3 }TensorTour Inc. \vspace{0.2em} \\
{\small \texttt{\{kranti, neermat, grv\}@cse.iitk.ac.in, siddharthsrivastava@cdac.in}}
}

\maketitle

\section{Introduction}
Humans exploit the tiny differences in the sound waves reaching both ears in terms of time and amplitude known as Interaural Time Difference (ITD) and Interaural Level Difference (ILD) to infer the spatial properties (eg. position of the sound source) from the sound. However, an audio recorded with a single microphone loses all these spatial properties present in the original sound. On the other hand, by using binaural audio we can recreate the original sound more accurately and thus giving the listener a feeling of being present in the recording place.

The aim of this work is to tackle the problem of audio binauralization, where the system takes a mono channel audio as input and predicts the corresponding two channel binaural audio.
In almost all the prior approaches on audio binauralization \cite{gao20192, lu2019self}, the visual information in the form of RGB image along with the audio input is used to predict the binaural audio. Although the appearance of the sound producing sources and their relative location in the scene serve as one of the important signal for the task, other important information like the distance of the source from the microphone or the geometry of the scene are completely ignored in these formulations. In some of the recent works \cite{richard2020neural, gebru2021implicit} similar information in the form of explicit position and orientation of the source and receiver are fed along with the audio input which improved the performance of the system as compared to using RGB images only. The downside with these approaches is that they require specialized equipment to track the position of both source and listener, which might not be feasible all the time. We address this by using depth features of the scene along with the visual appearance features as auxiliary signals in the process of audio binauralization. Further, image, depth and binaural audio have also been shown to be interrelated in \cite{parida2021beyond}. However, unlike \cite{parida2021beyond}, where the authors used binaural echoes to improve the depth prediction, here we perform the reverse task of using the depth features to obtain binaural audio.

As opposed to the prior approaches \cite{gao20192, zhou2020sep}, we use visual transformer \cite{ranftl2021vision} instead of convolutional layers as the backbone for extracting visual features. Deriving from the motivation that different audio components present in the sound should correspond to the location of sound producing objects and their depth in the scene we also propose a carefully tailored cross-modal attention to better associate the components of different modalities. 

\vspace{-0.5em}
\section{Approach}
\vspace{-0.5em}
\begin{figure*}
    \centering
    \includegraphics[scale=0.31]{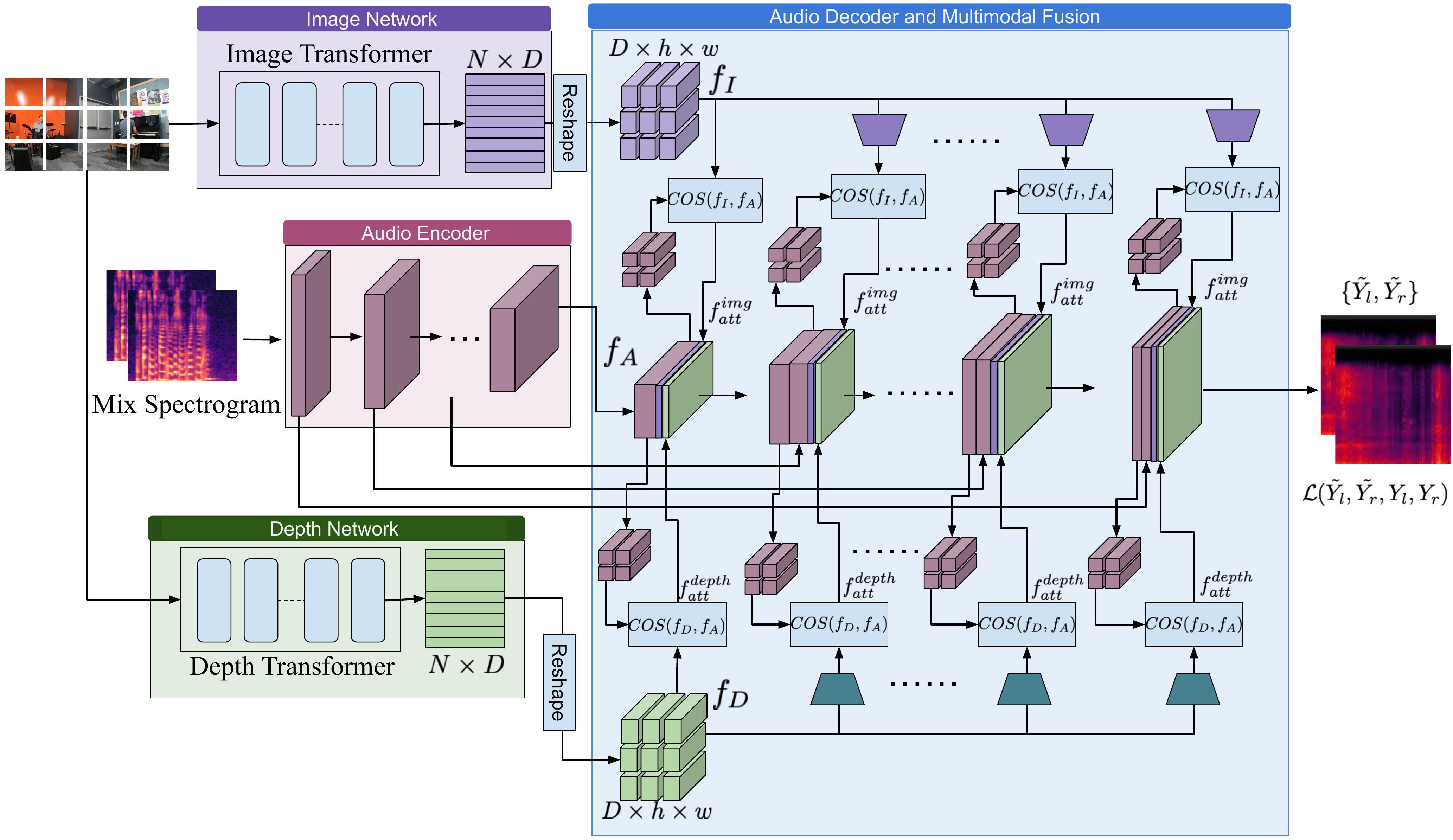}
    \caption{\textbf{Block diagram of proposed architecture. }The network inputs mono audio and RGB frame to produce corresponding binaural audio consistent with the visual scene. Image and Depth network inputs the same RGB frame producing image and depth features,  $f_\I$ and $f_\D$ respectively. Similarly, audio encoder inputs mono audio producing audio features $f_\A$. The image and depth features are then fed individually to a neural network to align the channels with audio features at each decoder layer. Finally, cross modal attention is obtained for both image-audio and depth-audio at each layer and then concatenated with audio features to obtain final predicted binaural audio.} \vspace{-1 em}
    \label{fig:block_diag}
\end{figure*}

Our task is to convert a mono channel audio, $x(t)$ to a binaural audio with $(y_l(t), y_r(t))$ as the corresponding left and right channels. To achieve this, we design a transformer network based deep neural network with three input modalities, i.e., an RGB image, depth features and a mono channel audio. The motivation behind using depth features comes from the amplitude difference between left and right channels in a binaural audio which correspond to the distance of the sound producing source. The visual and depth inputs enable the network to learn the association between binaural audio, depth of different sound producing sources and their relative position in the scene. To achieve this, we propose a network consisting of carefully tailored cross-modal attention mechanism to associate features from RGB, mono channel audio and depth.

Overall, as shown in Fig.\ref{fig:block_diag}, the network consists of three input blocks (i) audio network (ii) image network, (iii) depth network. The audio network is an encoder decoder architecture, whereas the image and depth network are self-attention transformer networks. We further perform cross-modal attention both for image-audio and depth-audio features separately at each layer of the decoder in audio network. We now describe each component below.

\subsection{Image and Depth Network}
\vspace{-0.5em}
For extracting image and depth features from an input RGB image, $\I \in \R^{3 \times H \times W}$, we use the recently proposed vision transformer (ViT) \cite{dosovitskiy2020image} backbone. We use ViT-Large architecture consisting of $24$ attention blocks for both image and depth features. Following \cite{ranftl2021vision}, we obtain features from four different layers of the transformer network, \ie $l \in \{6, 12, 18, 24\}$, to get information at the varying level of details. Finally to align the number of channels of depth, image and audio networks, we perform a $1 \times 1$ convolution on features from each of the four layers resulting in $d$ channels. We then concatenate features from all the four layers and obtain image and depth features $f_{\I} \in \R^{4d \times h \times w}$ and $f_{\D} \in \R^{4d \times h \times w}$ respectively. We now use both the features as input for hierarchical attention calculation in the network. We initialize the image and depth network with ImageNet \cite{deng2009imagenet} and MIX6 \cite{ranftl2021vision} respectively.

\subsection{Audio Encoder Network}
\vspace{-0.5em}
Similar to \cite{gao20192}, our audio encoder network consists of a UNet \cite{ronneberger2015u} style encoder-decoder architecture. We convert the time domain audio signal $x(t)$ into a STFT representation and concatenate both the real and imaginary to be fed as input to audio network, \ie $\A \in \R^{2 \times F \times T}$, where $F$ and $T$ are the no.\ of frequency bins and time steps in STFT respectively. It is then passed through successive layers of convolutions. Finally we obtain audio features, $f_{\A} \in \R^{d_A \times f \times t}$ as the output of audio encoder.

\subsection{Audio Decoder and Multimodal Fusion}
\vspace{-0.5em}
\noindent\textbf{Audio Decoder.} We adapt and build upon the audio decoder proposed in \cite{gao20192}. The decoder consists of $5$ fractionally strided convolutional layers, which increases the dimension of input tensor successively at each layer. The decoder network takes the input from all three modalities and produces a mask, $\M \in \mathbb R^{2 \times F \times T}$, with values in the range $[-1, 1]$. The final output is the difference between the right and left channel audio, $\Tilde{\O} \in \mathbb{R}^{2\times F\times T}$, predicted by multiplying the mask, $\M$ with the mixed input signal, $\A$, \ie $\Tilde{\O} = \M \cdot \A$. Finally, the left and right channel output in the STFT representation can be computed as $\Tilde{\Yl} = (\A + \Tilde{\O})/2$ and $\Tilde{\Yr} = (\A - \Tilde{\O})/2$ respectively.

\noindent\textbf{Multimodal Fusion.} The fusion operation combines the information from all three inputs at different scales. The image and depth features are already extracted from different layers. For audio, each time-frequency bin in the feature representation can be considered as an unique audio concept. Each of the audio concepts can act as the basic building blocks of different audio sources present in the scene and appropriate combination of these concepts can give back the original signal. These audio concepts represent different characteristics sound of the source. This characteristic sound can contain frequency variation within the source and a single audio concept can also contribute to multiple sound sources. e.g.\ the audio concept for a $7$ string \textit{guitar} can be the sound produced by each of the strings. Similarly an audio concept for \textit{acoustic guitar} can also be shared by \textit{electric or classical guitar} or by any similar sounding object such as \textit{piano} or \textit{saxophone}.

So, our goal in multimodal fusion is to effectively associate  different audio concepts to different object regions. Similar to the time-frequency bin in the audio representation, each co-ordinate in the spatial domain of the visual/depth features corresponds to certain region in the image. If the region contains a sounding object then the corresponding audio component should be weighted and also the depth value in the region should be used for the final output. We calculate two attention maps (i) between the image and audio features, and (ii) between the depth and audio features.

We design the network such that the output channels of audio encoder network is equal to the output channels of image and depth network, i.e.\ $d_I = d_D = d_A = 4d$. The attention is calculated between every pair of points.
{
\small
\begin{align}
    f_{att}^{img}(i,j,k, l) &= \frac{f_\I(:,i,j)^Tf_\A(:,k,l)}{\sqrt{\lVert f_\I(:,i,j) \rVert_2^2}\sqrt{\lVert f_\A(:,k,l) \rVert_2^2}} \forall i,j,k,l \\
    f_{att}^{depth}(i,j,k, l) &= \frac{f_\D(:,i,j)^Tf_\A(:,k,l)}{\sqrt{\lVert f_\D(:,i,j) \rVert_2^2}\sqrt{\lVert f_\A(:,k,l) \rVert_2^2}} \forall i,j,k,l
\end{align}
}
where, $f_{att}^{img}, f_{att}^{depth} \in \R^{h \times w \times f \times t}$ are the image-audio and depth-audio attention respectively. We then resize the 4D attention tensors into a 3D tensors such that the resulting attention maps are of size $[(h \times w) \times f_i \times t_i]$, where $f_i, t_i$ are the input spatial feature dimension in the $i^{th}$ layer of decoder. For the first decoder layer $f_i, t_i$ is exactly equal to the dimension of the audio encoder output. After the resizing operation the number of channel dimension in the feature map corresponds to all the distinct regions in the image. We interpret this as the attention weight in all the regions of the image for the particular audio concept. We obtain the final attention map at $i^{th}$ decoder level by concatenating both of them over the spatial axis.
\begin{align}
    f_{att}^{i} = \text{Concat}(\text{Resize}(f_{att}^{img}, f_{att}^{depth}))
\end{align}

Next, we concatenate the attention map with the audio features and feed the result to the next layer of the decoder. We also use skip connection at each decoder layer and concatenate features form corresponding encoder layer except for the first layer of decoder.

Finally, we calculate attention at each level of the decoder. As the audio network increases the feature dimension in each level, it also increases the time-frequency bins in the feature representation and hence the finer details in the audio comes successively with each decoding layer. In order to account for the coarse to fine representation of the audio we add the image and depth features similar to first layer to obtain the attention map. This is also in line with the approach proposed in \cite{zhou2020sep}. To perform the attention calculation at each layer, the feature channel of the image and depth should align with the channels of audio features. To perform the alignment we use a one-layer neural network followed by GELU non-linearity to make the feature dimensions of both the modalities equal. For matching the channels of audio and image feature, the one-layer neural network used for every layer has weights of dimension $[d_I, d_i]$ and $[d_D, d_i]$ for image and depth features respectively. Please note that for first layer of decoder, i.e.\ $i=1$, we do not use one-layer network as the channels are already aligned.

\subsection{Loss Function and Training}
Following earlier works~\cite{gao20192, zhou2020sep}, we use an L2 loss between the ground truth and network output given as
\begin{math}
    \mathcal{L}(\Tilde{\Yl}, \Tilde{\Yr}, \Yl, \Yr) = \lVert \Tilde{\Yl} - \Yl \rVert_2^2 + \lVert \Tilde{\Yr} - \Yr \rVert_2^2
\end{math} where $\Yl, \Yr$ are ground truth and $\Tilde{\Yl}, \Tilde{\Yr}$ are predicted left and right audio signals respectively. We finally train the whole network in an end-to-end manner.
\vspace{-0.5em}
\section{Experiments}
\vspace{-0.5em}
\noindent\textbf{Dataset.} We use the FAIR-Play dataset \cite{gao20192} for our experiments. We follow the same settings for input data representation, data preprocessing, distance metrics (STFT, Envelope (ENV) distance) and evaluation protocols as in earlier works \cite{gao20192, zhou2020sep}.

\noindent\textbf{Impact of adding depth.} To analyse the effectiveness of depth to the task of audio spatialization, we study the impact of  each modality on the performance of the network. We add each of the modality (image, depth) one by one to the network and then combine all the modalities to verify the contribution of each modality. For a fair comparison, we use exactly the same transformer architecture with equal number of parameters for both image and depth. The results are shown in Tab.\ref{tab:ablation}.

We report the performance in the \textit{split-1} of FAIR-PLAY dataset for all the models.
\begin{table}
    \centering
    \begin{tabular}{c|c|c}
    \hline
    Modality & STFT ($\downarrow$) & ENV ($\downarrow$) \\
    \hline \hline
    audio & 1.154 & 0.153 \\
    \hline
    audio+image & 0.942 & 0.141 \\
    audio+depth & 0.981 & 0.146 \\
    audio+image+depth& \textbf{0.928}& \textbf{0.140} \\
    \hline
    \end{tabular}
    \caption{\textbf{Audio binauralization by combining different modalities.} Using audio only (audio), audio with image features (audio+image), audio with depth features (audio+depth) and combination of audio, image and depth features (audio+image+depth). $\downarrow$ indicates lower is better. 
    }
    \label{tab:ablation}
    \vspace{-1 em}
\end{table}

From Tab.\ \ref{tab:ablation}, we observe improvement of $~18\%$ (STFT) and $~8\%$ (ENV) with image + mono audio as input over an audio only input (Tab \ref{tab:ablation}, row 1 vs row 2). Similarly we observe a decrease of $~15\%$ and $~5\%$ in STFT and ENV metrics respectively by using depth + mono audio as input as compared to audio input only (Tab \ref{tab:ablation}, row 1 vs row 3). This shows that both the image and depth features are helpful towards a better audio binauralization. As both the image and depth backbone contain exactly same number of parameters, the improvement in performance can be attributed to the information encoded in it. Adding image results in a better STFT over mono audio only input as compared to depth. This could be owed to the presence of semantic information in the RGB images in the form of appearance and relative location of different sound producing regions. However, as depth input has relative distance information within the scene, and results in better binauralization as compared to mono audio input, we hypothesize that combining depth with RGB will provide more contextual information leading to better localization by the network. This is also evident with the empirical performance of adding both depth and image features along with audio, which results in an improvement of $~20\%$ and $~9\%$ in STFT and ENV respectively over mono audio input (Tab \ref{tab:ablation}, row 1 vs row 4).

\noindent\textbf{Comparison to state-of-the-art.} The comparison is given in Tab.\ref{tab:sota}. APNet \cite{zhou2020sep} has two additions over Mono2Binaural \cite{gao20192}. First, it performs multi-task optimization by combining the task of audio separation along with audio binauralization instead of a single task. Second, it uses two losses to train the network, one with the difference (Diff.) of both the output as in \cite{gao20192} and another with direct prediction of audios from both the channels (2 Ch.). We report the performance for all the combinations of loss and tasks in Tab.\ \ref{tab:sota}. We can observe that the proposed method obtains an improvement of $~22\%$ over the baseline of audio only, $~6.4\%$ over Mono2Binaural \cite{gao20192}, $7.9\%$ over APNet (Multi Task, 2 Ch. loss) and $~3.75\%$ over APNet (Multi task, Diff. loss). Further, we also observe that our method slightly lags behind the variants of APNet using both the losses with or without multi task optimization. However, it is important to note that the multi task optimization in APNet uses additional training data, which is not required by our method.

\begin{table}
    \centering
 \resizebox{\columnwidth}{!}{
    \begin{tabular}{c|c|c}
    \hline
    Modality & STFT ($\downarrow$) & ENV ($\downarrow$) \\
    \hline \hline
    audio & 1.154 & 0.153\\
    \hline
    Mono2Binaural \cite{gao20192} & 0.959 & 0.141\\
    APNet (Multi Task, 2 Ch. loss) \cite{zhou2020sep}& 0.976 & 0.145\\
    APNet (Multi Task, Diff. loss) \cite{zhou2020sep}& 0.933 & 0.140\\
    \hline
    Ours (Diff.) & 0.898 & 0.138 \\
    \hline
    APNet (2Ch. + Diff. loss) \cite{zhou2020sep}& 0.889 & 0.136 \\
    APNet (Multi Task, 2 Ch.+Diff. loss) \cite{zhou2020sep}& 0.879 & 0.135 \\
    \hline
    \end{tabular}
     }
    \caption{\textbf{Comparison with existing approaches} We report the results for existing approaches directly from \cite{zhou2020sep}. $\downarrow$ indicates lower is better. 
    }
    \label{tab:sota}
    \vspace{-1.5 em}
\end{table}

\begin{figure}
    \centering
    \includegraphics[width=0.99\columnwidth]{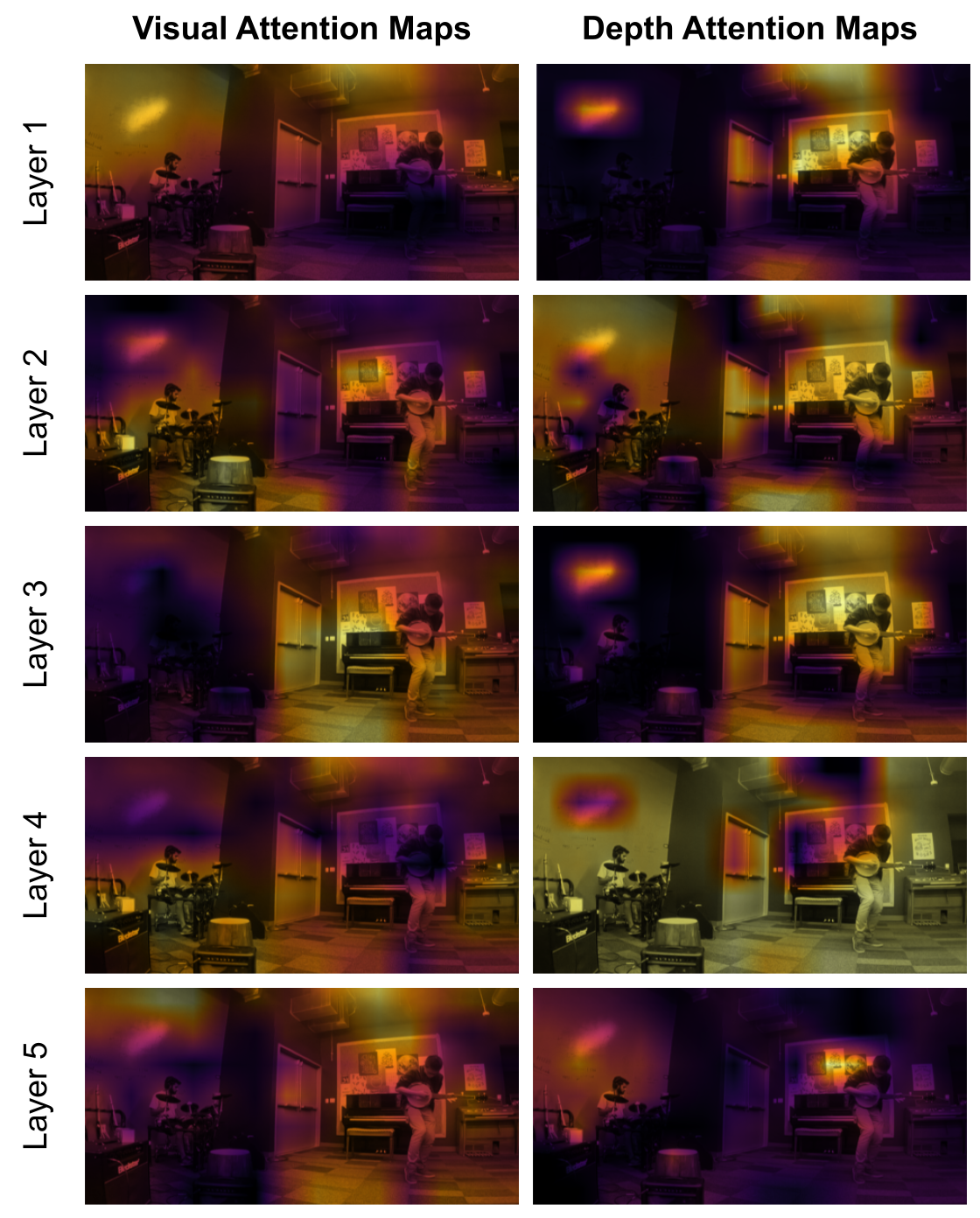}
    \caption{\textbf{Attention Map Visualization} Attention maps for both visual and depth channel at each decoder layer. We observe that visual attention map progressively attends to the sound producing regions in the image where as the depth attention maps attends to the structure of the room, i.e.\ wall, ceiling and floor.} 
    \label{fig:qual_viz}
    \vspace{-1em}
\end{figure}

\noindent\textbf{Qualitative Results.} We give qualitative results of visual and depth attention map obtained from all layers of decoder in Fig.\ \ref{fig:qual_viz}. From the visual attention map in first column, we observe that the attention values are spread out over the entire image in the first layer but in successive layers of 2,3, and 4 it produces high values only to the sound sources. We also observe that layer 3 produces high values for the source on the right side of the image whereas layer 4 produces high values for the source in the left side of the image. This region specific attention map can be considered as the inherent association between left and right audio channel with left and right regions of the image, which is important for an effective binauralization. For the depth attention maps, we observe that instead of attending to the sound source location, it looks at different structure of the rooms such as wall, ceiling and floor in layer 1, 2, and 3 of the decoder. From these attention maps, we make a general observation that the depth network infuses information about the geometry of the room resulting in better binauralization. 

\vspace{-1em}
\section{Conclusion}
\vspace{-0.5em}
We proposed an end-to-end trainable multi-modal transformer network for mono to binaural audio generation. We studied the impact of image, depth and audio only input along with their combinations on this task. We noticed that adding depth provides additional structural information which assists in better source localization as visually analysed from attention maps.

\vspace{-0.5em}
{\small
\bibliographystyle{ieee}
\bibliography{reference}
}

\end{document}